 \documentclass[final,5p,times,twocolumn,authoryear]{elsarticle}

\usepackage{amssymb}
\usepackage{lipsum}
\usepackage{subcaption}
\usepackage{verbatim}

\usepackage[dvipsnames]{xcolor} 

\usepackage{ulem}  

\journal{Physics Letters B}

\begin{document}

\begin{frontmatter}


\ead{mleize@df.uba.ar}

\title{Tensions in cosmology: a discussion of statistical tools to determine inconsistencies}


\author[UBA,IFIBA-CONICET]{Matías Leizerovich}
\author[UBA,IFIBA-CONICET]{Susana J. Landau}
\author[UNLP, CONICET]{Claudia G. Scóccola}
\affiliation[UBA]{organization={Universidad de Buenos Aires, Facultad de Ciencias Exactas y Naturales, Departamento de Física.},
            addressline={Ciudad Universitaria, 1428}, 
            city={Capital Federal},
            postcode={1460}, 
            state={Buenos Aires},
            country={Argentina}}
\affiliation[IFIBA-CONICET]{organization={CONICET - Universidad de Buenos Aires, Instituto de Física de Buenos Aires (IFIBA).},
            addressline={Ciudad Universitaria, 1428}, 
            city={Capital Federal},
            postcode={1460}, 
            state={Buenos Aires},
            country={Argentina}}
\affiliation[UNLP]{organization={Facultad de Ciencias Astronómicas y Geofísicas, Universidad Nacional de La Plata, Observatorio Astronómico},
            addressline={Paseo del Bosque, B1900FWA}, 
            city={La Plata},
            postcode={}, 
            state={Buenos Aires},
            country={Argentina}}
\affiliation[CONICET]{organization={Consejo Nacional de Investigaciones Científicas y Técnicas (CONICET)},
            addressline={Godoy Cruz 2290}, 
            city={Capital Federal},
            postcode={1425}, 
            state={Buenos Aires},
            country={Argentina}}

\begin{abstract}
We present a comprehensive analysis of statistical tools for evaluating tensions in cosmological parameter estimates arising from distinct datasets. Focusing on the unresolved Hubble constant ($H_0$) tension, we explore the Pantheon Plus + SH0ES (PPS) compilation, which includes low-redshift Cepheid data from the SH0ES collaboration, along with the latest release of CMB data from the Planck collaboration, Cosmic Chronometers (CC) dataset and the most recent Baryonic Acoustic Oscillation (BAO) datasets. Employing various tension metrics, we quantitatively assess the inconsistencies in parameter estimates, emphasizing the importance of capturing multidimensional tensions. Our results reveal substantial tension between PPS and Planck 2018 datasets and moderate tension between the BAO data sets and all other datasets. We highlight the importance of adopting these metrics to enhance the precision of future cosmological analyses and facilitate the resolution of existing tensions.
\end{abstract}

\begin{keyword}

cosmology \sep statistical tools \sep tension metrics \sep  Pantheon Plus + SH0ES \sep Cosmic Chronometers \sep BAO \sep Planck data

\end{keyword}

\end{frontmatter}

\section{Introduction} \label{introduction}

In the past twenty years, the amount and precision of cosmological data has increased significantly. As a result, stringent constraints were established for the cosmological parameters such as the baryon and dark matter density parameters, the Hubble constant, and others. On the other hand, it is well known that the predictions of the standard cosmological model are in agreement with the majority of the observations. However, there are tensions between the values of some of the cosmological parameters obtained with different datasets. These discrepancies remain an open question for cosmologists and are usually used as a motivation for studying alternative cosmological models. One of the most important unsolved issues is the so-called $H_{0}$ tension: the value of the current Hubble parameter $H_{0}$ that has been obtained using data from the Cosmic Microwave Background (CMB) assuming a standard cosmological model \citep{Planckcosmo2018} is not in agreement with the one inferred in a model-independent way, with data from type Ia supernovae  and Cepheids \citep{Riess2022}. This issue has been extensively discussed in the literature but there is no agreement within the community about the source of the discrepancy \citep{2021ApJ...919...16F,2023arXiv230810954R,2022PhR...984....1S,Vagnozzi_2023}.
Another issue is the one called $S_8$ tension, namely, the difference between the estimation of this parameter (defined as $S_8=\sigma_8 \left(\frac{\Omega_m}{3}\right)^{0.5}$) from CMB and BAO data with the one obtained from weak lensing and galaxy clustering data \citep{Kids,DES3year}. Hence, the ability to quantify differences in the estimation of cosmological parameters when different datasets are considered is essential for the success of present and future research in Cosmology. 

Given two posterior distributions obtained from two different datasets A and B, the most common method to study statistical tension is to evaluate the one-dimensional marginalization of the posterior distributions. 
A \textit{rule of thumb} formula is introduced in~\cite{2021MNRAS.505.6179L} to measure the discrepancy, expressed in terms of the number of standard deviations $\sigma$, between parameter estimations $\theta$ derived from distinct datasets. The formulation is as follows:
\begin{equation}
    N_{\theta} = \frac{|\mu_{A}-\mu_{B}|}{\sqrt{\sigma_{A}^{2}+\sigma_{B}^{2}}}\, ,
    \label{eq: rule_of_thumb}
\end{equation}
where the means $\mu_{A/B}$ and the variances $\sigma_{A/B}$ correspond to the ones of the posteriors obtained from each dataset. However, this method carries some problems. For instance, marginalization can hide tensions that can only be seen in higher dimensions. This is caused by the fact that marginalization over some of the parameters necessarily implies a loss of information. Moreover, the number of dimensions of the problem also affects the inferred tension. The significance of the discrepancy in the parameter estimations from the two experiments depends on the number of shared parameters effectively constrained by both experiments. Therefore, a variety of methods have been developed to determine the consistency between the posterior distributions that are obtained from different datasets. For example, different tension metrics that quantify this problem in the whole parameter space at once have been studied in \cite{PhysRevD.99.043506} and \cite{PhysRevD.104.043504}. 
In this work, we focus on a particular type of metrics, namely the ones that are based on the posterior distributions, to study tensions between different data sets\footnote{In \cite{2021MNRAS.505.6179L} these are called \textit{parameter-space methods}.}. We analyze the tension in the most recently released Pantheon Plus + SH0ES (PPS) \citep{Scolnic:2021amr} with three different datasets: the CMB Planck 2018 release (Planck18), Baryonic Acoustic Oscillation (BAO) BOSS and eBOSS most recent datasets, and Cosmic Chronometers data compilation (CC). Although the tension between PPS and Planck18 datasets may be familiar to the reader, the tension between CC and PPS has not been studied in detail, despite being mentioned in \cite{Moresco:2023zys}. Since CC data only contains information about the Hubble parameter, a non-trivial tension in the comparison between PPS and CC may point to a tension in the Hubble constant at the background level, providing new insights into the study of the $H_{0}$ tension.

This work is organized as follows: in Section \ref{sec:tension_metrics} we describe the tension metrics that we apply to the different datasets, while in Section \ref{sec:datasets} we present the results of the statistical analyses that show inconsistencies in the obtained confidence intervals of some parameters. In Section \ref{sec:results} we show the amount of inconsistency that we obtain applying the method proposed in this article, that is, from the metrics described in Section \ref{sec:tension_metrics} and compare it with the one inferred from the rule of thumb. We also discuss these results in light of the interpretation of what each metric is quantifying. Finally, in Section \ref{sec:conclusion} we present the conclusions of our work.

\section{Tension metrics} \label{sec:tension_metrics}
In this section we describe the different metrics that have been introduced in \cite{PhysRevD.99.043506,PhysRevD.104.043504}. We recall the Bayes formula that defines the posterior distribution 
\begin{equation}
    P(\theta|D)=\frac{\mathcal{L}(D|\theta)\cdot\Pi(\theta)}{\varepsilon(D)}\, ,
\end{equation}
where $\mathcal{L}(D|\theta)$ is the likelihood of the data D given the parameters $\theta$, $\Pi(\theta)$ is the prior distribution and $\varepsilon(D)=\int \mathcal{L}(D|\theta)\cdot\Pi(\theta) \, d\theta$ is the Bayesian Evidence\footnote{This expression also depends on the assumed physical model $\mathcal{M}$, which is fixed in our analysis.}. We can divide the metrics into two types \citep{2021MNRAS.505.6179L}: the ones based on Bayesian Evidence $\varepsilon(D)$ and the ones based on the Posterior distributions $P(\theta|D)$. In this work we focus on the last group, which was implemented in the \textit{Tensiometer} repository\footnote{https://github.com/mraveri/tensiometer}. All the results are presented in terms of an effective number of standard deviation, $N_{\sigma}$, which is defined by \citep{PhysRevD.99.043506}
\begin{equation}
    {\rm P} = {\rm Erf}(N_{\sigma}/\sqrt{2})
    \label{eq: N_sigma}
\end{equation}
where $P$ corresponds to a probability that will identify agreement or disagreement between the considered datasets while $N_{\sigma}$ is the number of standard deviations associated to a 1D gaussian distribution with probability $P$. Moreover, the interpretation of the probability $P$ varies with the chosen metric and will be clarified for each case in what follows.

\subsection{Gaussian metrics}

\begin{figure*}
    \centering 
    \includegraphics[width=0.7\textwidth]{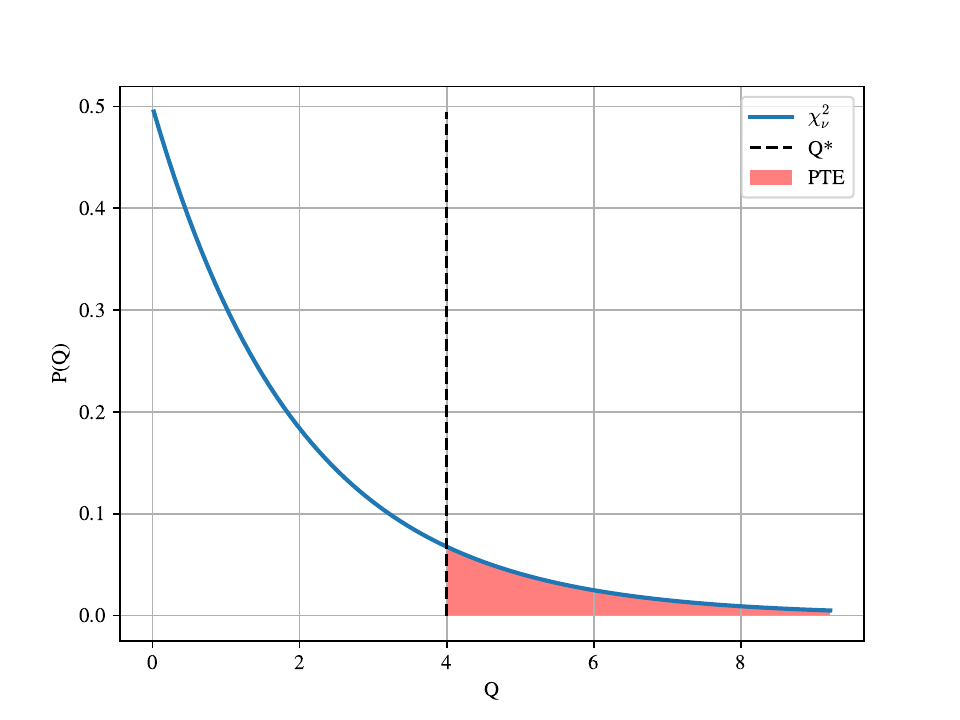}
    \caption{\textit{An example of a quadratic estimator probability density distribution. In red, it is shown the area that represents the probability to exceed, PTE.}}
    \label{fig:PTE}
\end{figure*}

Firstly, we describe tension metrics associated to quadratic estimators, which assumed gaussian posterior distributions. In Fig. \ref{fig:PTE} we show the area that corresponds to the probability to exceed (PTE) certain observed value of the estimator Q, which is noted as $Q*$.  The assigned probability to estimate the tension is $P=1-\rm{PTE}$, and the equivalent $N_\sigma$ is obtained using eq.~\ref{eq: N_sigma}.

\subsubsection{Parameter differences in standard form}

This method consists in calculating the statistical difference between the inferred parameters given by dataset A $(\theta_{A})$ and the ones given by dataset B $(\theta_{B})$, with the estimator
\begin{equation}
    Q_{\rm DM}= (\hat{\theta}_{B}-\hat{\theta}_{A})^{T} (\hat{C}_{B}+\hat{C}_{A})^{-1}(\hat{\theta}_{B}-\hat{\theta}_{A})\, ,
    \label{eq: Q_DM}
\end{equation}
where $(\hat{\theta}_{i})$ and $\hat{C}_{i}$ are the mean and the covariance matrix on the parameter space obtained from the statistical analysis with dataset $i$ (where $i$ correspond to dataset A or B). If both posterior distributions are gaussian, then $Q_{\rm DM}$ follows a $\chi^{2}_{\nu}$ distribution with $\nu=\rm{Rank}[\hat{C}_{B}+\hat{C}_{A}]$ degrees of freedom \footnote{The rank of a matrix A is equal to the dimension of the vector space generated by the columns of A.}. We stress that this metric measures the difference in the obtained confidence intervals of all parameters, while also including the effect of correlations between them. Therefore, it can be regarded as a reasonable generalization of the rule of thumb, with the additional benefit that it quantifies all inconsistencies jointly.

\subsubsection{Parameter differences in updated form}

Firstly introduced in \cite{PhysRevD.99.043506}, this method consists in calculating the statistical difference between the inferred parameters given the dataset A $(\theta_{A})$ and the ones given by the joint datasets $(\theta_{AB})$, with the estimator
\begin{equation}
    Q_{\rm UDM}= (\hat{\theta}_{AB}-\hat{\theta}_{A})^{T} (\hat{C}_{AB}-\hat{C}_{A})^{-1}(\hat{\theta}_{AB}-\hat{\theta}_{A})\, ,
    \label{eq: Q_UDM}
\end{equation}
where $\hat{C}_{AB}$ is the covariance matrix obtained from the statistical analysis with the joint datasets. It can be seen that if the inferred parameters for both datasets A and A+B are gaussian distributed, $Q_{\rm UDM}$ has a $\chi^{2}_{\nu}$ distribution with $\rm \nu=rank[(\hat{C}_{AB}-\hat{C}_{A})]$ degrees of freedom. 
As it was shown in \cite{PhysRevD.99.043506}, in order to compute $Q_{\rm UDM}$ it is convenient to break the posterior distribution as a sum over the Karhunen–Loéve (KL) modes of the covariances involved. With this approach, the $Q_{\rm UDM}$ estimator has a $\chi^{2}$ distribution with $\nu$ degrees of freedom, where $\nu$ is the number of the KL eigenvalues that are taken into account. We can point out two interesting features of this metric. Firstly, the directions of the parameter space that shows significant tension can be identified \textit{a priori}, which helps its physical interpretation. Secondly, non-gaussianities are mitigated since we can select the most constraining directions in the parameter space by the two datasets. Finally, it is worth noticing that this metric is asymmetric, since eq.~\ref{eq: Q_UDM} is not invariant under the exchange of $\rm A\rightarrow B$. We stress that this metric answers a different question than the other ones, namely, how the results of the statistical analysis using a given dataset are updated if a new dataset is added. For this reason, in our view it is not recommended to quantify the tension with UDM metric if a strong tension has been detected with the other metrics, for example, DM.

\subsubsection{Goodness-of-fit loss}
This estimator measures the difference in the likelihood function evaluated at the maximum values of the posterior distribution, considering the two datasets jointly and separately in the statistical analysis
\begin{equation}
    Q_{\rm DMAP} = 2 ln \mathcal{L}_{A}(\theta_{pA}) + 2 ln \mathcal{L}_{B}(\theta_{pB}) - 2 ln \mathcal{L}_{A+B}(\theta_{pA+B})\, ,
    \label{eq: Q_DMAP}
\end{equation}
where $\theta^{A/B}_{p}$ are the Maximum a posteriori (MAP) parameters considering the dataset A/B. Notice that this statistic does not depend explicitly on the covariance of the distribution and only compares the evaluation of the likelihood on certain points. In the case in which the likelihood and the posterior are gaussian distributed, the estimator $Q_{\rm DMAP}$ has a distribution $\chi^{2}$ with $\Delta \nu$ degrees of freedom, where
\begin{equation}
    \Delta \nu = \nu^{A} + \nu^{B} - \nu^{A+B}\, .
\end{equation}
In this case, the number of degrees of freedom is defined as 
\begin{equation}
    \nu = N - tr[C_{\Pi}^{-1}\,C_{p}] \, ,
\end{equation}
where $N$ is the number of data points and $C_{\Pi}$, $C_{p}$ are the covariance matrix of the Prior and Posterior distributions, respectively. We emphasize that this metric provides a good quantification of the difference between how well the theoretical predictions can describe both data sets jointly with respect to the same situation but considering the two data sets separately.
Therefore, it is a nice tool regarding how the data fit the theoretical prediction but does not give a good estimate of the inconsistency between parameters.

\subsection{Non gaussian metrics: Exact Parameter Shift}

This method is based on the computation of the parameter difference probability density $P(\Delta\theta)$, where $\Delta\theta=\theta_{A}-\theta_{B}$ is the difference between the means of the posterior parameters that correspond to datasets $A/B$. The general expression for two uncorrelated datasets is given by \citep{PhysRevD.99.043506,PhysRevD.104.043504}
\begin{equation}
    P(\Delta\theta) = \int P_{A}(\theta)P_{B}(\theta-\Delta\theta)\,d \theta\, ,
    \label{eq: Delta_theta}
\end{equation}
which is analogous to the expression of the cross-correlation function in signal processing. The statistical significance of the shift is calculated by summing over all the values of $P(\Delta\theta)$ over the isocontour corresponding to no shift $\Delta\theta = 0$:
\begin{equation}
    \Delta = \int_{P(\Delta\theta)>P(0)} P(\Delta\theta)\,d\Delta\theta.
    \label{eq: Delta}
\end{equation}
Here the assigned probability to identify the tension is $P=\Delta$  and the equivalent $N_\sigma$ is calculated inverting eq.~\ref{eq: N_sigma}.
Although it may look like the evaluation of the last equation is straightforward, it is particularly difficult when the parameter space is high dimensional. Taking the difference between the samples of the two MCMC chains as it was described in \cite{PhysRevD.101.103527}, we can generate the chain of the parameter difference, which is an estimation of the convolution integral in eq.~\ref{eq: Delta_theta}. Finally, the integration in eq.~\ref{eq: Delta} can be accomplished using Kernel Density Estimation (KDE) methods. This method does not assume a particular form of the posterior distribution. 
This metric measures the discrepancy directly from the chains of the parameter difference defined above. Therefore, like $Q_{DM}$, it quantifies the tension directly from the outputs of the MCMC process. However, in the case that the isocontour of $P(\Delta \theta =0)$ is far from the maximum of $P(\Delta \theta)$, the output of the metric is difficult to compute.

Next, we point out some final observations about the metrics described before. As it was shown in eq.~\ref{eq: Q_DM}, DM metric quantifies the difference between the parameters obtained from datasets A and B separately, weighted by the sum of their covariance matrices as it was shown in eq.~\ref{eq: Q_DMAP}. Furthermore, DMAP metric quantifies the difference between the fit with the joint datasets (A+B) with respect to the ones using the datasets separately. Consequently, both metrics quantify different aspects of the tension in the parameter space and it is not trivial to compare between them.
 On the other hand,  $Q_{UDM}$ evaluates how the results using one dataset are updated when including another one. In short, in the case of gaussian posteriors, $Q_{DM}$ and Parameter Shift quantify the discrepancy between inferred 
parameters directly from the outputs of the inference process and therefore should be considered the best metrics to quantify discrepancies between parameters.

\section{Parameter inferences with different datasets} \label{sec:datasets}

In this paper, we compute the estimators described in Section \ref{sec:tension_metrics} for different combinations of four cosmological datasets. Firstly, we consider the recently released Pantheon Plus compilation of type Ia supernovae (SNIa). We note that this release includes the option of using low redshift  Cepheid data obtained by the SHOES collaboration, which are crucial for the calibration of SNIa and therefore for the Hubble tension. Therefore, we name this data set as Pantheon Plus + SHOES (PPS). The Pantheon Plus compilation consists of 1,701 SNIa at redshifts between $0.0012 < z < 2.26$, which are available in \cite{Scolnic:2021amr}\footnote{https://github.com/PantheonPlusSH0ES/DataRelease}. Secondly, we consider data obtained with the Cosmic Chronometers technique detailed in \cite{simon05,stern10,moresco12,zhang14,10.1093/mnrasl/slv037,CC2}. Thirdly, we consider the most recent BOSS and eBOSS datasets \cite{10.1093/mnras/stu2693, 10.1093/mnras/stx721, PhysRevD.103.083533, 10.1093/mnras/staa2800, 10.1093/mnras/staa2455, du_Mas_des_Bourboux_2020, 10.1093/mnras/staa3234, 10.1093/mnras/staa2780}. Finally, we use CMB data from the latest release of Planck \citep{Planckcosmo2018} (Planck18).

To estimate the posterior distribution we use the latest available version of \textit{CLASS} \citep{DiegoBlas_2011} and perform a Markov Chain Monte Carlo (MCMC) analysis with the latest release of the \textit{MontePython} software \citep{BRINCKMANN2019100260,Audren:2012wb}. Table \ref{Tab: priors} shows the priors on the shared parameter space\footnote{In Planck18, $\Omega_{m}$ and $H_{0}$ are derived parameters, while in BAO the derived parameter is $\omega_{b}$.} that are considered in  our analyses. Note that the prior in the physical baryon density $\omega_{b}=\Omega_{b} h^{2}$ is only used in the comparison between Planck18 and BAO.

\begin{table}
\centering
\begin{tabular}{l c c c}
\hline
Parameter    &  $\Omega_{m}$  &  $H_{0}$     &  $\omega_{b}$  \\
\hline \hline
Prior range  &  [0.1 , 0.9]   &  [20 , 100]  &  [$4\,\times 10^{-4}$ , 0.3]  \\
\hline \hline
\end{tabular}
\caption{\textit{Flat priors on the three cosmological parameters of the standard model used for all analyses in this work. $\Omega_{m}$ and $H_{0}$ are derived parameters in Planck18, while the same applies to $\omega_{b}$ in BAO.}}
\label{Tab: priors}
\end{table}

\begin{figure*}
\centering 
    \begin{subfigure}{0.49\textwidth}
        \includegraphics[width=\textwidth]{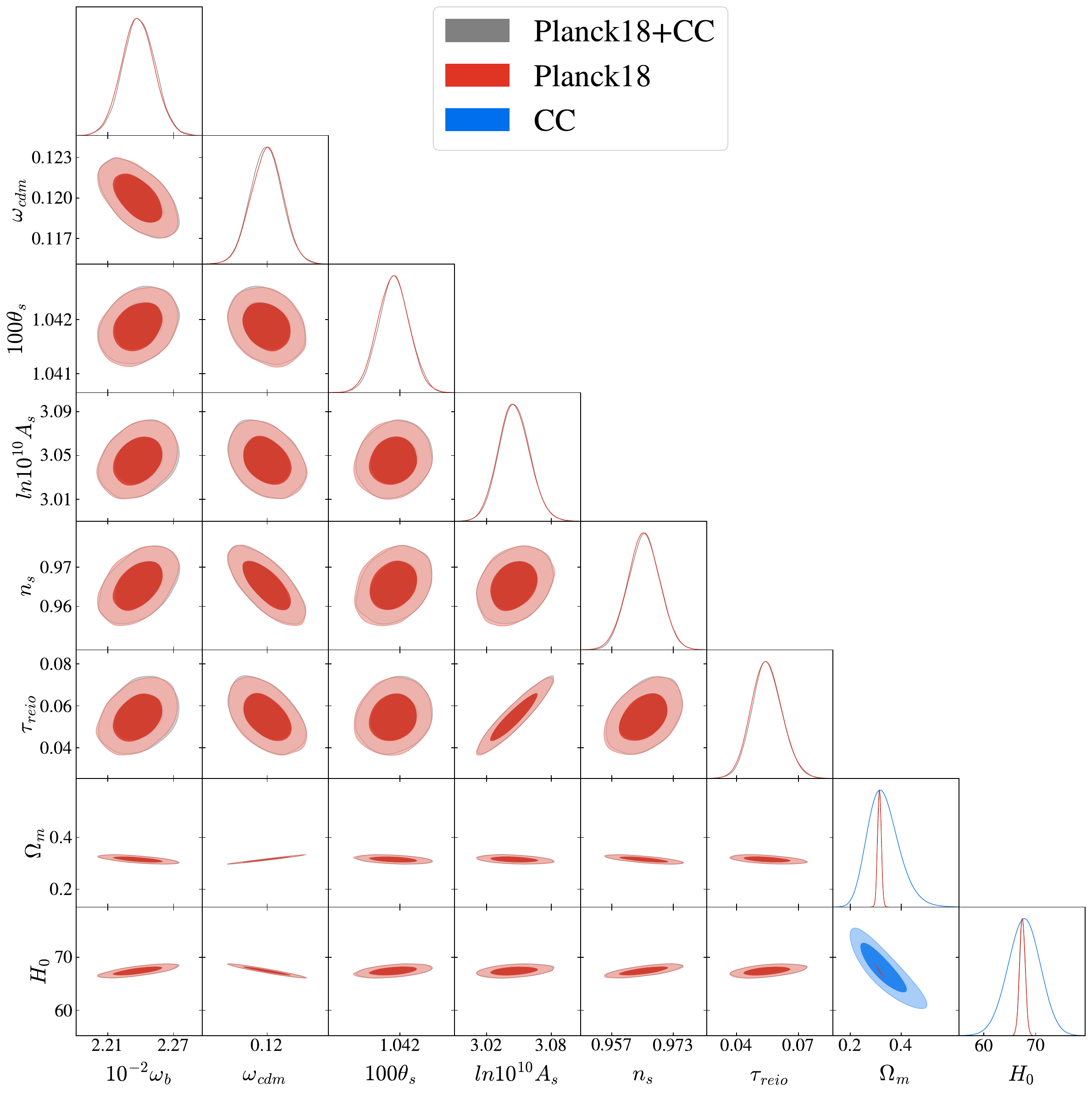}
    \end{subfigure}
    \hfill
    \begin{subfigure}{0.49\textwidth}
        \includegraphics[width=\textwidth]{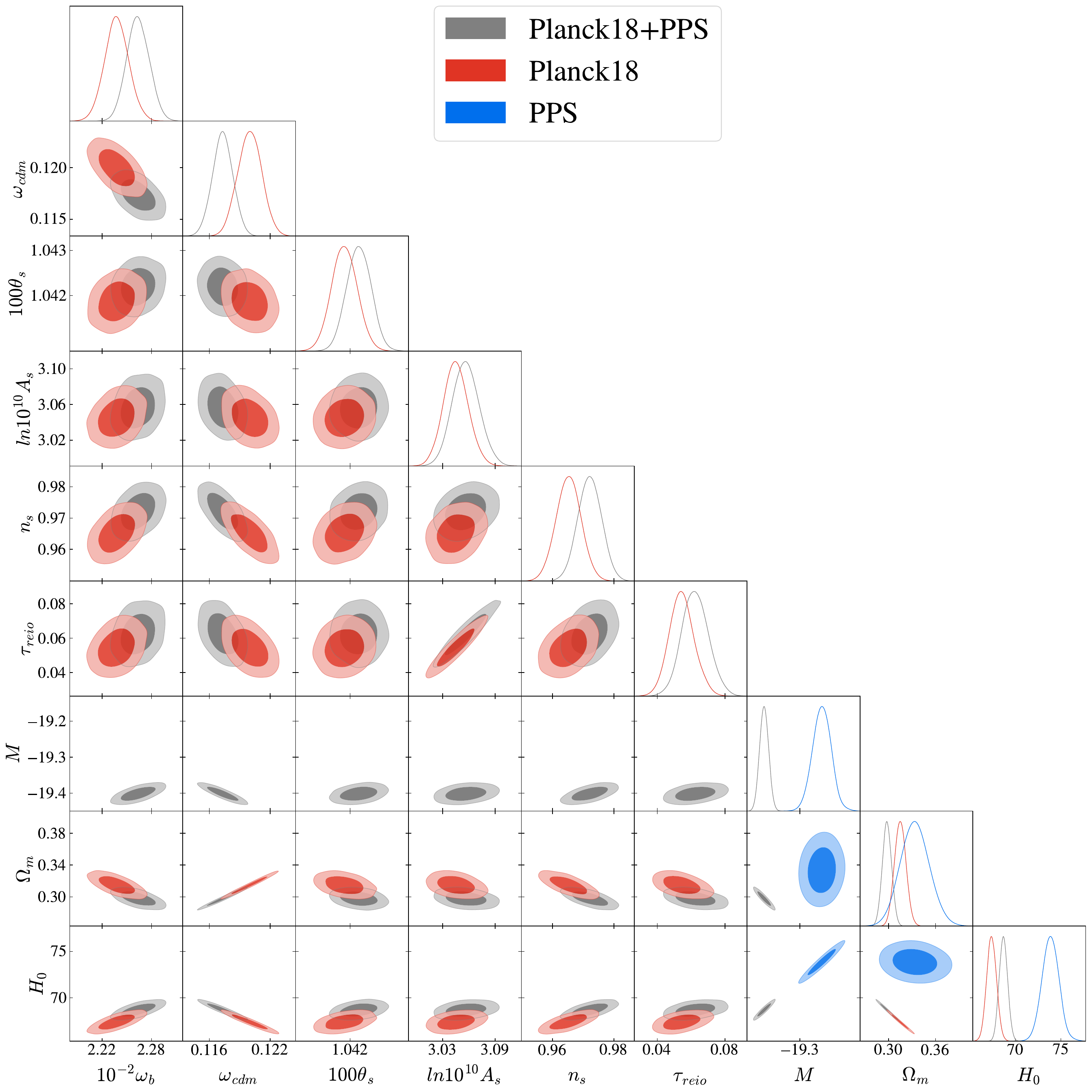}
    \end{subfigure}
    \caption{\textit{Results of the statistical analysis for the $\Lambda \rm{CDM}$ model using  Planck likelihood 2018 with Cosmic Chronometers (left)} and with Pantheon Plus + SH0ES (right), respectively. The darker and brighter regions correspond to 65\% and 95\% confidence regions, respectively. The plots in the diagonal show the posterior probability density for each of the free parameters of the model.}
    \label{fig:contours_full}
\end{figure*}

\begin{figure*}
    \centering 
    \includegraphics[width=0.5\textwidth]{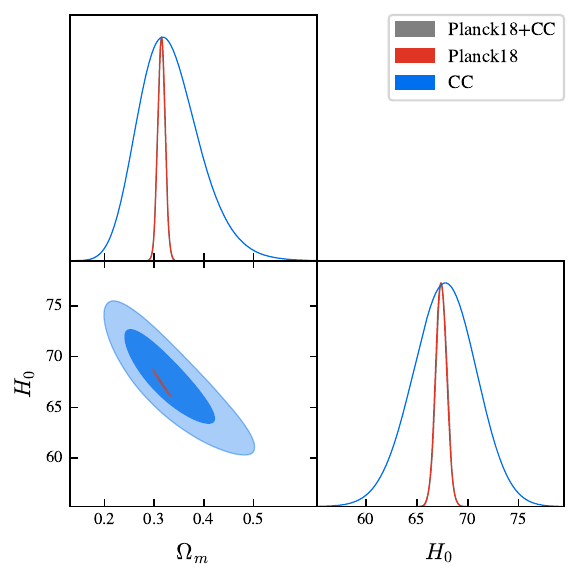}
    \caption{Results of the statistical analysis for the $\Lambda \rm{CDM}$ showing the marginalization on the shared plane on the parameter space $\Omega_{m}-H_{0}$ for Planck18+CC. The darker and brighter regions correspond to 65\% and 95\% confidence regions, respectively. The plots in the diagonal show the posterior probability density for each of the free parameters of the model.}
    \label{fig:contours_Planck18_CC}
\end{figure*}

\begin{figure*}
    \centering 
    \begin{subfigure}{0.49\textwidth}
            \includegraphics[width=\textwidth]{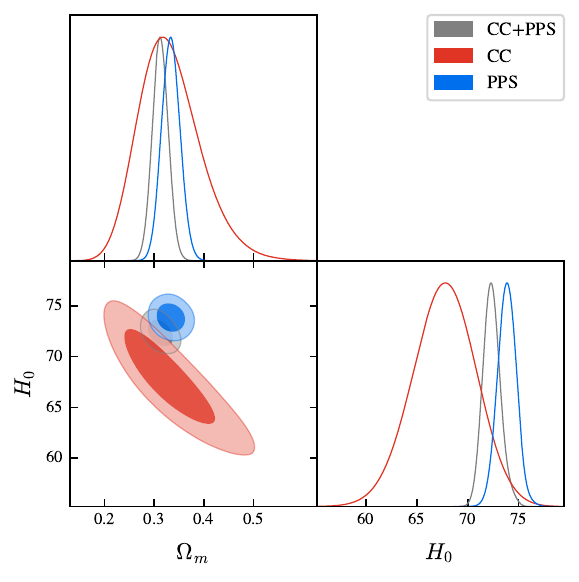}
        \end{subfigure}
        \hfill
        \begin{subfigure}{0.49\textwidth}
            \includegraphics[width=\textwidth]{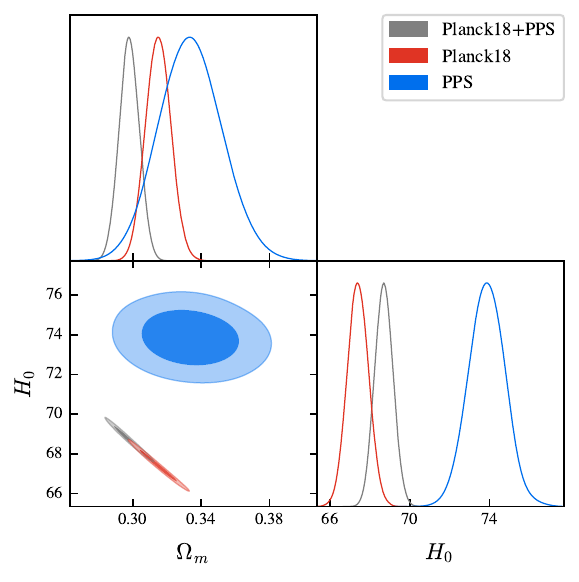}
        \end{subfigure}
    \caption{Results of the statistical analysis for the $\Lambda \rm{CDM}$ showing the marginalization on the shared plane on the parameter space $\Omega_{m}-H_{0}$. The darker and brighter regions correspond to 65\% and 95\% confidence regions, respectively. The plots in the diagonal show the posterior probability density for each of the free parameters of the model.}
    \label{fig:contours_PPS}
\end{figure*}

\begin{figure*}
    \centering 
    \begin{subfigure}{0.49\textwidth}
            \includegraphics[width=\textwidth]{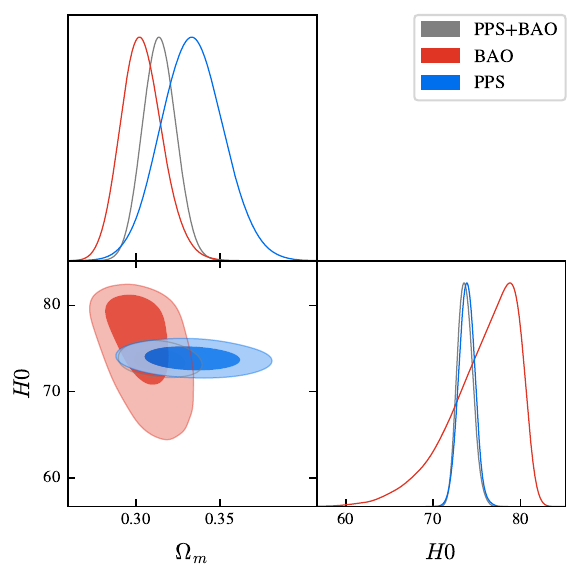}
        \end{subfigure}
        \hfill
        \begin{subfigure}{0.49\textwidth}
            \includegraphics[width=\textwidth]{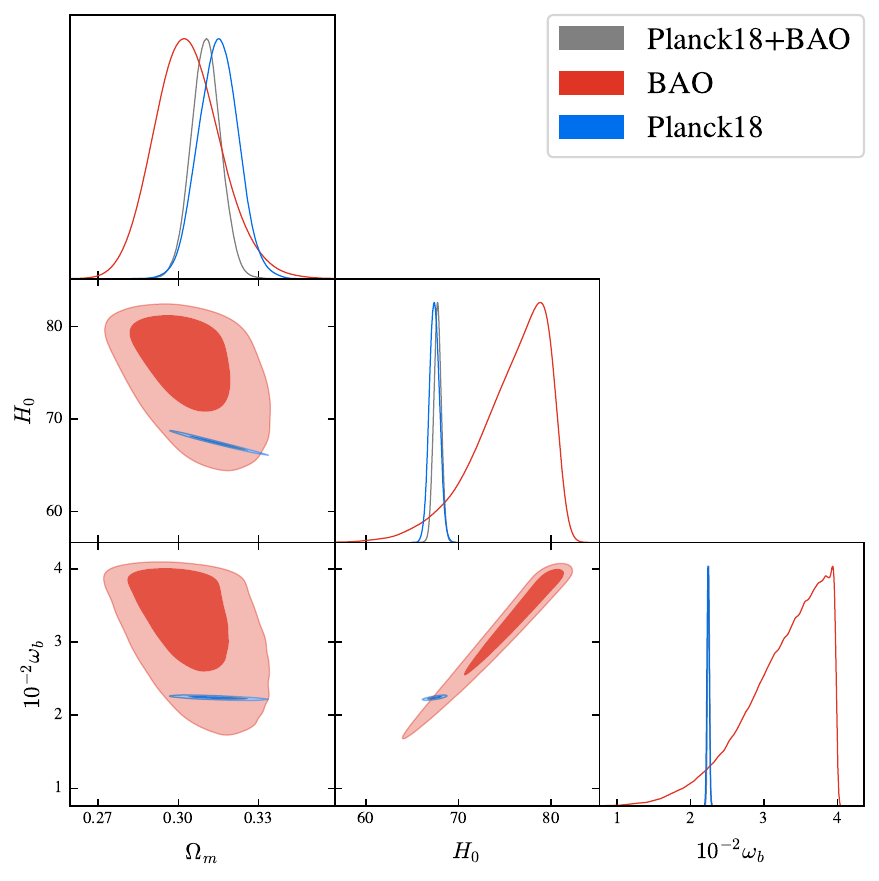}
        \end{subfigure}
    \caption{Results of the statistical analysis for the $\Lambda \rm{CDM}$ showing the marginalization on the shared plane on the parameter space $\Omega_{m}-H_{0}$ for BAO+PPS and $(\Omega_{m}, H_{0}, \omega_{b})$ for Planck18+BAO. The darker and brighter regions correspond to 65\% and 95\% confidence regions, respectively. The plots in the diagonal show the posterior probability density for each of the free parameters of the model.}
    \label{fig:contours_BAO}
\end{figure*}

For the CC dataset, the matter density $\Omega_{m}$ and the Hubble parameter $H_0$ are the free parameters of the model. For PPS, the Supernovae absolute magnitude $M_{abs}$ has to be taken into account as a nuisance parameter. For BAO, we consider $(\Omega{}_{m },\, H_{0},\, \Omega{}_{b })$. Finally, when Planck18 data is taken into account, all parameters of the standard cosmological model, ${\rm \Lambda CDM}$,   $(\omega{}_{b },\, \omega{}_{cdm },\, \theta{}_{s },\, A_{s },\, n_{s },\, \tau{}_{reio })$ are considered, apart from the nuisance parameters of the Planck likelihood. In this case, $\Omega_{m}$ and $H_0$ can be obtained as derived parameters of this analysis. Our analysis is focused on the comparison of different pairs of the mentioned datasets. Fig.~\ref{fig:contours_full} shows the results of the statistical analysis performed considering Planck18/CC and Planck18/PPS datasets jointly and separately\footnote{In the case of Planck18, we do not show the contours corresponding to the nuisance parameters of the Planck likelihood.}.
It follows that for Planck18/CC (left) the datasets are in agreement, while for Planck18/PPS the tension is considerable large. Besides, it is worth noticing that the shared parameter space for all datasets is the $\Omega_{m}-H_{0}$ plane, except for the pair Planck18/BAO in which the shared parameter space is $(\Omega_{m},\, H_{0}, \, \omega_{b})$.  The projections of the posterior distributions on the shared parameter subspace are shown in Figs. \ref{fig:contours_Planck18_CC}, \ref{fig:contours_PPS}, and \ref{fig:contours_BAO}. Finally, we point out that the posteriors are gaussian distributions in all cases. This is because priors are not informative.

\section{Results} \label{sec:results}

Here we present the results of our analysis. The pair of datasets that have been compared are Planck/CC, CC/PPS, PPS/BAO, Planck18/PPS and Planck18/BAO. Before presenting the results for each pair of datasets, we discuss the results in general for all the metrics.

\subsection{Discussion for each metric}\label{subsec:metrics_results}

\subsubsection{DM}

Here we discuss DM metric in view of all our results. According to eq. \ref{eq: Q_DM}, the tension for this metric is proportional to the difference between the means of the posterior distributions, and to the inverse of the sum of the covariance matrices. Following this definition, we encounter two paradigmatic cases: Firstly, the tension between Planck18/CC presents the lowest tension of all the analyses (as it is shown in Fig. \ref{fig:contours_Planck18_CC}, the difference between the means of each posterior is negligible). Secondly, the tension between Planck18/PPS is the strongest one (the difference between the means is the most important one, as it is shown in Fig. \ref{fig:contours_PPS} (right)). Besides, it is remarkable that DM metric indicates a different level of tension with respect to the sum of the 1D rule of thumb for all analysis. This is reasonable since the rule of thumb does not take into account the correlation between parameters.

Finally, we encountered three cases with moderate tension: CC/PPS, Planck18/BAO and PPS/BAO \footnote{This last case shows weak to moderate tension}. Let us briefly discuss the first two cases. Although in the 1D projection it seems that CC/PPS presents weaker tension than Planck18/BAO, the 2D projection indicates that it is the opposite case. The fact that DM metric takes into account the covariance of the posterior distributions allows to quantify better the tension between datasets in moderate cases.

\subsubsection{UDM}
According to eq. \ref{eq: Q_UDM}, the tension for this metric is proportional to the difference between the means of the posterior distributions of one of the datasets and the joint analysis, and to the inverse of the difference between their covariance matrices. As it was discussed before, this metric is not symmetric and we expect different results when some dataset is updated with another one. In order to gain intuition of UDM results, we analyze two relevant cases.

Firstly, we analyze the case of CC/PPS. When CC results are updated with PPS, the obtained tension is $\sim 2.4 \sigma$, while when PPS is updated with CC the tension is $\sim 4.8 \sigma$. This can be explained as follows. According to eq.~\ref{eq: Q_UDM}, the estimator may indicate higher tension if the difference in the means is high or if the difference $|C_{AB}-C_{A}|$ is small. In this case, the covariance matrix of the posterior corresponding to PPS and PPS+CC are quite similar which makes the estimator $Q^*_{UDM}$ and the corresponding $N_\sigma$ higher.

Secondly, we analyze the case of Planck18/PPS. When Planck18 results are updated with PPS the equivalent $N_\sigma$ is $\sim 2.3 \sigma$, while when PPS is updated with Planck18 we obtain $\sim 4.6 \sigma$. Contrary to the case of PPS+CC, here the dominant effect on the metric is the difference between the means of the compared distributions, rather than the difference in the covariance matrices, as it was shown in the right panel of Fig. \ref{fig:contours_PPS}. We note that in this case, the UDM metric has one effective degree of freedom. This is because the KL decomposition points out that the majority of the variance is condensed in the first eigenvalue.

\subsubsection{DMAP}

According to eq. \ref{eq: Q_DMAP}, DMAP compares the likelihoods of the two datasets separately with the one of the joint analysis, all of them evaluated at the maximum of their corresponding posterior distributions. If both datasets are independent, we can write eq. \ref{eq: Q_DMAP} as

\begin{equation}
    Q_{\rm DMAP} = 2 \ln{ \left( \frac{\mathcal{L}_{A}(\theta_{pA}) \cdot \mathcal{L}_{B}(\theta_{pB})}{\mathcal{L}_{A}(\theta_{pA+B}) \cdot \mathcal{L}_{B}(\theta_{pA+B})}\right)}  \, .
\end{equation}

This equation shows that the comparison is between the separate likelihoods evaluated at the maxima of their posteriors and the same likelihoods evaluated at the maxima of the joint posterior distribution. Assuming gaussianity, when $\theta_{pA+B}$ is far from $\theta_{pA}$ or $\theta_{pB}$, this results in a tension on the DMAP metric.

Unlike DM, this metric does not take into account the covariance matrix of the distributions but has information on the joint analysis. For this reason, both DM and DMAP quantify the tension in different ways and their results cannot be compared. However, like DM, this metric shows that the strongest tension also corresponds to Planck18/PPS and the weakest to Planck18/CC. Finally, since we are using flat uninformative priors, the effective number of degrees of freedom for all analyses is equal to the dimensions of the shared parameter space in all cases.

Also, we discuss two cases of moderate tension: CC/PPS and PPS/BAO. To facilitate the discussion, we  assume  that the   posterior distribution  is equal to the corresponding likelihood and that no relevant information is lost in the marginalization process\footnote{Even in the case of uninformative priors, prior volume effects in the marginalization process can lead to a loss of information, as has been discussed in \cite{Gomez-Valent:2022hkb}. To solve this issue, these authors suggest the use of profile distributions.  However, the latter work also shows that for the $\Lambda$CDM model and data sets similar to the ones used here, our assumption is valid.}. The left panel of Fig. \ref{fig:contours_PPS} (CC/PPS) shows that the maximum of the joint posterior is located in the $2\sigma$ contour of the corresponding posteriors obtained considering only one dataset, and the left panel of Fig. \ref{fig:contours_BAO} (PPS/BAO) shows a similar behaviour but this time the maximum of the joint posterior is placed in the $1\sigma$ contour of the corresponding contours obtained with only one dataset. This is in agreement with the results shown in Tables \ref{Tab: comparison_PPS_CC} and \ref{Tab: comparison_PPS_BAO}.

\subsubsection{Exact Parameter Shift}
\begin{figure*}
    \centering 
    \begin{subfigure}{0.3\textwidth}
        \includegraphics[width=\textwidth]{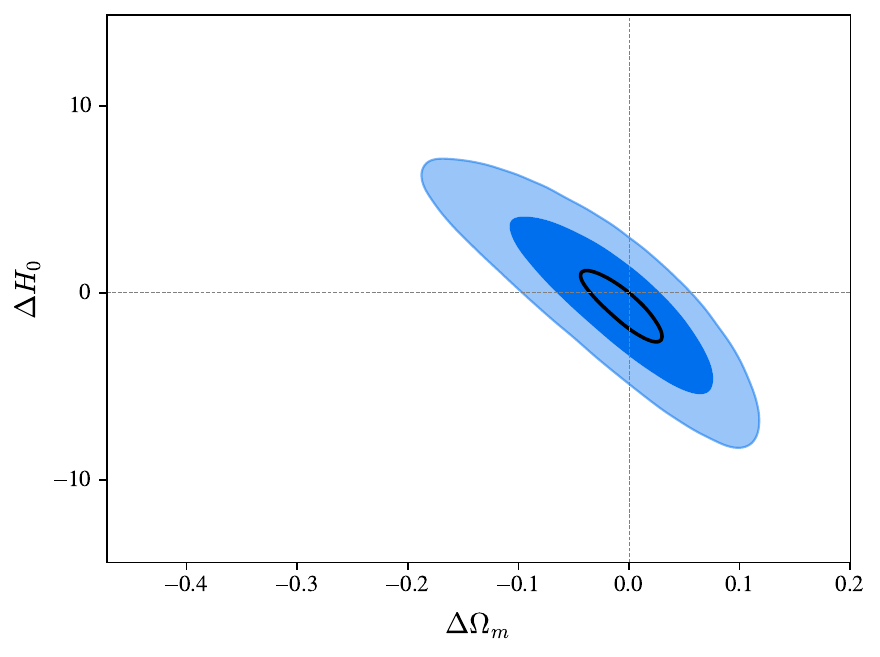}
    \end{subfigure}
    \hfill
    \begin{subfigure}{0.3\textwidth}
        \includegraphics[width=\textwidth]{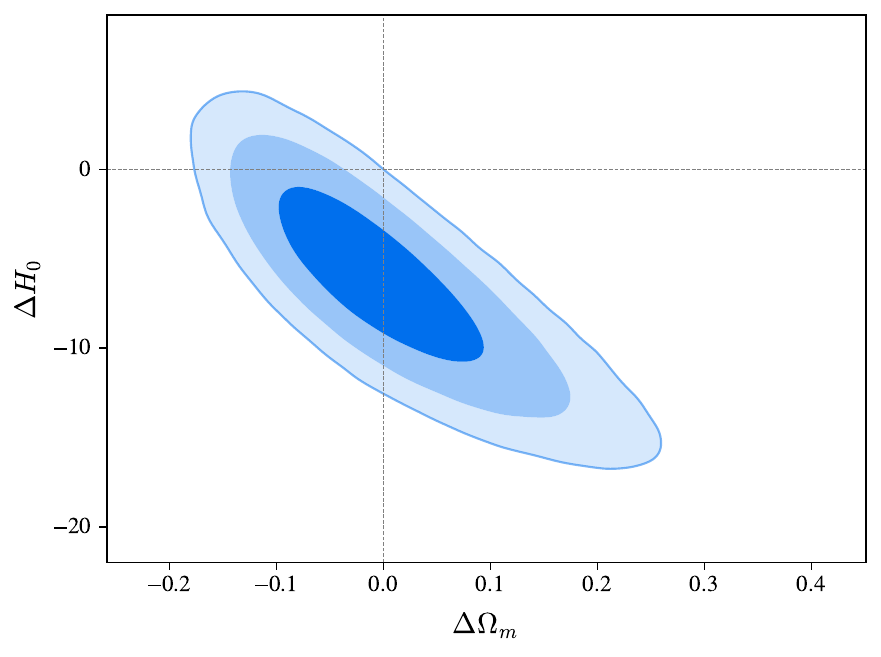}
    \end{subfigure}
    \hfill
    \begin{subfigure}{0.3\textwidth}
        \includegraphics[width=\textwidth]{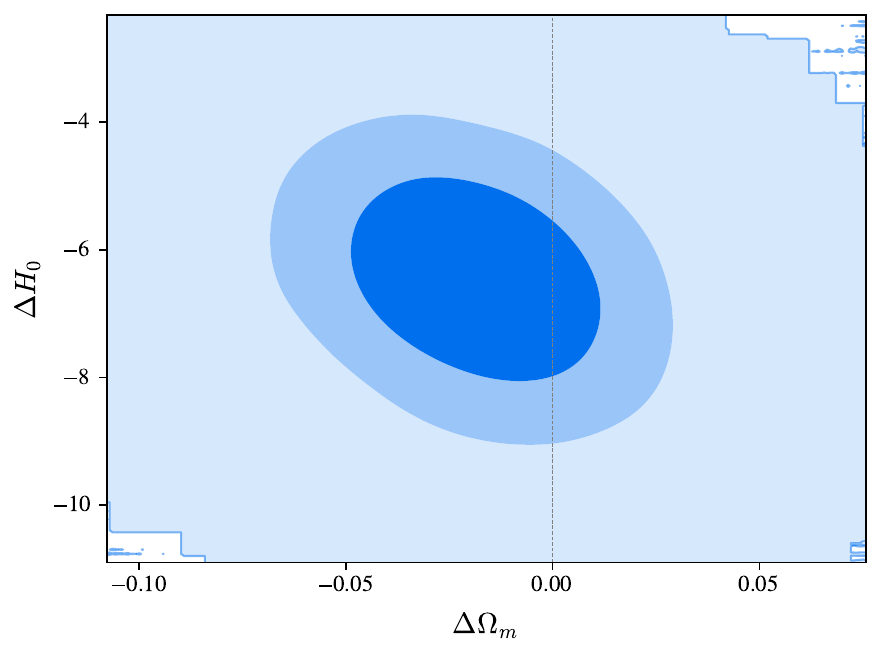}
    \end{subfigure}
    \caption{Distribution of parameter differences defined in eq.~\ref{eq: Delta_theta}, for the datasets Planck18/CC (left), CC/PPS (center) and Planck18/PPS (right). The first two delimited contours correspond to 65\%, 95\% confidence region. On the first figure, the black contour represents the contour that corresponds to $P(\Delta\theta=0)$. On the last two figures, this contours is outside the 95\% confidence region.}
    \label{fig:contours-diff}
\end{figure*}

Here we discuss our results for Exact Parameter Shift. In Fig. \ref{fig:contours-diff}, we show the distribution of $P(\Delta \theta)$ for three different levels of tension: the Planck18/CC comparison with negligible tension (left), the CC/PPS comparison with moderate tension (center) and the Planck18/PPS comparison with strong tension (right). Apart from the $1\sigma$ and $2\sigma$ contours, the last two cases show the isocontour that corresponds to $P(\Delta\theta=0)$. 

Besides, we expect that when the distributions are gaussian, the results of this metric is similar to the one of $Q_{DM}$. Indeed, this is the case for CC/PPS, Planck18/BAO, Planck18/CC and PPS/BAO. It is particularly relevant the tension between Planck18/PPS, in which Exact Parameter Shift gives an infinite tension. Although this result may be uncomfortable, it can be explained as follows: there is no isocontour of $P(\Delta\theta)$ that corresponds to $P(\Delta\theta=0)$ as it was shown in Fig. \ref{fig:contours-diff} (right). According to eqs. \ref{eq: N_sigma} and \ref{eq: Delta}, this implies that $\Delta=1$ and so the probability is translate to an infinite number of $\sigma$. We also checked that this effect is not due to a problem of sampling.

\subsection{Comparison between pairs of datasets}

In what follows, we describe our results for the tension metrics when taking the different datasets in pairs.

\subsubsection{Comparison between Planck18 and CC}

The comparison between Planck18 and CC shows weak tension (see Table \ref{Tab: comparison_Planck_CC} and Fig. \ref{fig:contours_Planck18_CC}). As we discussed in Subsec. \ref{subsec:metrics_results}, in this comparison we expect maximum consistency between datasets. Is is worth noting that in all cases, the tension is higher than the one reported by the rule of thumb. This example shows that even in the case of no full consistency, the tension is higher when all shared parameter space is considered.

\begin{table}
\centering
\begin{tabular}{l c c c c}
\hline
                                                 & $\nu$   & Q*       & PTE       & \textit{$N_{\sigma}$}  \\
\hline \hline
DM                                               &    2    &  0.54   &  0.76      &  0.300                  \\
\hline
$\rm UDM_{Planck18\rightarrow Planck18+CC}$      &    2    &  1.06   &  0.59      &  0.541                  \\
\hline
$\rm UDM_{CC\rightarrow  Planck18+CC}$           &    1    &  0.12   &  0.73      &  0.348                  \\
\hline
DMAP                                             &    2    &  0.63   &  0.73      &  0.345                  \\
\hline
Exact Param. Shift                               &    -    &  -      &  0.83      &  0.212                 \\

\hline
\hline
\multicolumn{3}{l}{Rule of thumb}                                    & $N_{\Omega_{m}}$ = 0.280 \\
\hline
\multicolumn{3}{l}{}                                                 & $N_{H_{0}}$      = 0.0995  \\
\hline \hline
\end{tabular}
\caption{Results of the application of different metrics on the datasets of Planck18/CC. The results are presented in terms of the probability P for a $\chi^2_\nu$ distribution with $\nu$ degrees of freedom when it corresponds (or $1-\Delta$ in the case of Exact Parameter Shift) and in all cases for the number of standard deviations \textit{$N_{\sigma}$}.}
\label{Tab: comparison_Planck_CC}
\end{table}

\subsubsection{Comparison between PPS and CC}

The comparison between PPS and CC is shown in Table \ref{Tab: comparison_PPS_CC} and Fig. \ref{fig:contours_PPS} (left). As we discussed in Subsec. \ref{subsec:metrics_results}. DM and Exact Parameter Shift match their results as expected. Besides, DMAP metric indicates an smaller tension. Finally, note that DM and DMAP indicate moderate tension.

\begin{table}
\begin{tabular}{l c c c c}
\hline
                                                 & $\nu$       & Q*          & PTE                       & \textit{$N_{\sigma}$}      \\
\hline \hline
DM                                               &     2       &    10.38    &  $5.57\,\times 10^{-3}$   & 2.772                      \\
\hline
$\rm UDM_{CC \rightarrow CC+PPS}$                &     2       &    8.62     &  $1.35\,\times 10^{-2}$   & 2.471                      \\
\hline
$\rm UDM_{PPS \rightarrow CC+PPS}$               &     2       &    26.58    &   $1.35\,\times 10^{-4}$  & 4.778                      \\
\hline
DMAP                                             &     2       &    4.85     &   $8.84\,\times 10^{-2}$  & 1.704                      \\
\hline
Exact Param. Shift                               &     -       &      -      &   $4.80\,\times 10^{-3}$  & 2.820                      \\
\hline
\hline
\multicolumn{3}{l}{Rule of thumb}                                    & $N_{\Omega_{m}}$ = 0.0171  \\
\hline
\multicolumn{3}{l}{}                                             & $N_{H_{0}}$ = 1.899        \\
\hline \hline
\end{tabular}
\caption{Results of the application of different metrics on the datasets of CC/PPS. The results are presented in terms of the probability PTE for a $\chi^2_\nu$ distribution with $\nu$ degrees of freedom when it corresponds (or $1-\Delta$ in the case of Exact Parameter Shift) and in all cases for the number of standard deviations \textit{$N_{\sigma}$}.
}
\label{Tab: comparison_PPS_CC}
\end{table}

\subsubsection{Comparison between Planck18 and BAO}

Comparison between Planck18 and BAO is presented in Table \ref{Tab: comparison_Planck_BAO} and Fig. \ref{fig:contours_BAO} (right) and shows a weak to moderate tension. Unlike the previous analyses, in this case the shared parameter space has three dimensions: $(\Omega_{m}, H_{0}, \omega_b)$. This implies that we cannot visualize the total likelihood but only the 2D projections. As expected, DM metric and Exact Parameter Shift are in agreement, while DMAP shows a weaker tension than the other ones.

\begin{table}
\centering
\begin{tabular}{l c c c c}
\hline
                                                 & $\nu$   & Q*     & PTE     & \textit{$N_{\sigma}$}  \\
\hline \hline
DM                                               &    3    &  4.62  &  0.20  &  1.276                 \\
\hline
$\rm UDM_{Planck18\rightarrow Planck18+BAO}$     &    3    &  1.45  &  0.69   &  0.395                 \\
\hline
$\rm UDM_{BAO\rightarrow  Planck18+BAO}$         &    1    &  0.21  &  0.65   &  0.460                 \\
\hline
DMAP                                             &    3    &  3.28  &  0.35   &  0.933                 \\
\hline
Exact Param. Shift                               &    -    &    -   &  0.14  &  1.473                 \\

\hline
\hline
\multicolumn{3}{l}{Rule of thumb}                                    & $N_{\Omega_{m}}$ = 0.767 \\
\hline
\multicolumn{3}{l}{}                                                 & $N_{H_{0}}$      = 2.041 \\
\hline
\multicolumn{3}{l}{}                                                 & $N_{\omega_b}$   = 1.847 \\
\hline \hline
\end{tabular}
\caption{Results of the application of different metrics on the datasets of Planck18/BAO. The results are presented in terms of the probability P for a $\chi^2_\nu$ distribution with $\nu$ degrees of freedom when it corresponds (or $1-\Delta$ in the case of Exact Parameter Shift) and in all cases for the number of standard deviations \textit{$N_{\sigma}$}.}
\label{Tab: comparison_Planck_BAO}
\end{table}

\subsubsection{Comparison between PPS and BAO}

Comparison between PPS and BAO is shown in Table \ref{Tab: comparison_PPS_BAO} and Fig. \ref{fig:contours_BAO} (left). In this case, all metrics indicate a moderate tension. The results for the DM metric and Exact Parameter Shift are similar, while DMAP metric indicates a higher tension than the other ones.

\begin{table}
\centering
\begin{tabular}{l c c c c}
\hline
                                                 & $\nu$   & Q*      & PTE     & \textit{$N_{\sigma}$} \\
\hline \hline
DM                                               &    2    &  1.80   &  0.41    &  0.829                \\
\hline
$\rm UDM_{BAO\rightarrow PPS+BAO}$               &    2    &  2.32   &  0.314   &  1.007                \\
\hline
$\rm UDM_{PPS\rightarrow  PPS+BAO}$              &    2    &  2.52   &  0.284  &  1.072                \\
\hline
DMAP                                             &    2    &  1.55   &  0.460   &  0.739                \\
\hline
Exact Parameter Shift                            &    -    &  -      &  0.311   &  1.014                     \\

\hline
\hline
\multicolumn{3}{l}{Rule of thumb}                                    & $N_{\Omega_{m}}$ = 1.339 \\
\hline
\multicolumn{3}{l}{}                                                 & $N_{H_{0}}$      = 0.434  \\
\hline \hline
\end{tabular}
\caption{Results of the application of different metrics on the datasets of PPS/BAO. The results are presented in terms of the probability P for a $\chi^2_\nu$ distribution with $\nu$ degrees of freedom when it corresponds (or $1-\Delta$ in the case of Exact Parameter Shift) and in all cases for the number of standard deviations \textit{$N_{\sigma}$}.}
\label{Tab: comparison_PPS_BAO}
\end{table}

\subsubsection{Comparison between PPS and Planck18}

Finally, the comparison between PPS and Planck18 is shown in Table \ref{Tab: comparison_Planck_PPS} and Fig. \ref{fig:contours_PPS} (right). As we discussed in Subsec. \ref{subsec:metrics_results}, this is the case that shows the strongest tension. For the DM metric, we obtain a tension of $\sim 6.4 \sigma$, which is not far from the estimation obtained from the rule of thumb for $H_0$. The most relevant result of this analysis is the infinite number of standard deviations using the Exact Parameter Shift metric, which has been already discussed in Subsec. \ref{subsec:metrics_results}.

\begin{table}
\centering
\begin{tabular}{l c c c c}
\hline
                                                 & $\nu$   & Q*          & PTE                      & \textit{$N_{\sigma}$}       \\
\hline \hline
DM                                               &     2   &    45.44    &   $1.36\,\times 10^{-10}$  & 6.420                       \\
\hline
$\rm UDM_{Planck18\rightarrow Planck18+PPS}$     &     1   &    5.36     &   $2.06\,\times 10^{-2}$   & 2.316                       \\
\hline
$\rm UDM_{PPS\rightarrow  Planck18+PPS}$         &     1   &    21.34    &   $3.84\,\times 10^{-6}$   & 4.620                       \\
\hline
DMAP                                             &     2   &    21.23    &   $2.46\,\times 10^{-5}$   & 4.218                       \\
\hline
Exact Parameter Shift                               &     -   &      -      &      0                   & $\infty$                    \\

\hline
\hline
\multicolumn{3}{l}{Rule of thumb}                                    & $N_{\Omega_{m}}$ = 0.942   \\
\hline
\multicolumn{3}{l}{}                                                 & $N_{H_{0}}$ = 6.093        \\
\hline \hline
\end{tabular}
\caption{Results of the application of different metrics on the datasets of Planck18/PPS. The results are presented in terms of the probability P for a $\chi^2_\nu$ distribution with $\nu$ degrees of freedom when it corresponds (or $1-\Delta$ in the case of Exact Parameter Shift) and in all cases for the number of standard deviations \textit{$N_{\sigma}$}.}
\label{Tab: comparison_Planck_PPS}
\end{table}

\section{Summary and conclusions} \label{sec:conclusion}

In this work, we discuss the importance of using tension metrics to determine tensions between different datasets. We show that the different metrics analyzed here can quantify the tension more precisely than the widely applied Rule of Thumb.

Among the metrics used, three of them (i.e., DM, DMAP, Exact Parameter Shift) quantify the difference between the inferred parameters while the other one (UDM) measures how much one dataset updates the results of another when added to the statistical analysis. Also, we discuss some implementation details; for example, some computational difficulties appeared when computing the Exact Parameter Shift while analyzing the tension between Planck18 and PPS.

Our results show two extreme cases: i) a very good agreement between CC and Planck18, and ii) a strong tension between PPS and Planck18. We also find three intermediate cases with moderate tension (BAO and Planck18; BAO and CC; BAO and PPS). These moderate tensions  were not pointed out nor quantified before. Therefore,  our analyses show that the tension metrics are excellent tools for quantifying moderate tension between distinct data sets.
With the availability of new data in the near future, it is expected that the errors will be reduced, and the posterior contours will be narrower, potentially increasing the tension between datasets. 

Applying the tension metrics that we have analyzed here, allows us to perform a detailed and precise analysis of the tension between different datasets, including also the correlation between parameters. 
We emphasize that the use of tension metrics determines the tension between distinct datasets in the plane of shared parameters rather than the tension in a single parameter as in the case of the rule of thumb. 
We expect that the use of these metrics will become relevant for the analysis of future datasets.
For example, some of these metrics have been used to discuss the recently released DESI results~\citep{2024arXiv240403002D}. 
The increase in the precision of new results may lead to stronger tensions, which can be quantitatively described using tension metrics.

Finally, we discuss which metric is best to quantify the tension between datasets. To give a final answer, several aspects have to be taken into account. Firstly, as it has been pointed out before, UDM metric quantifies how a new dataset updates another one, and its results are not comparable with the ones of DM/Exact Parameter Shift or DMAP. Secondly, although the rest of the metrics indeed quantify the tension between independent datasets, DM/Exact Parameter Shift and DMAP quantify different aspects of the tension: while DMAP quantifies the distance between the MAP evaluating the likelihood of the joint posterior and the one using the datasets separately, DM quantifies the distance between the means of the independent posteriors weighted by their covariance matrix. Therefore, in our opinion, DM/Exact Parameter Shift answers accurately the question about the tension between independent datasets.

\section*{Acknowledgements}
S.L. and M.L. are supported by grant PIP 11220200100729CO CONICET and grant 20020170100129BA UBACYT. 
C.G.S. is supported by grant PIP-2876 CONICET, and grant G175 from UNLP.

\bibliographystyle{elsarticle-harv} 
\bibliography{references}

\begin{thebibliography}{33}
\expandafter\ifx\csname natexlab\endcsname\relax\def\natexlab#1{#1}\fi
\providecommand{\url}[1]{\texttt{#1}}
\providecommand{\href}[2]{#2}
\providecommand{\path}[1]{#1}
\providecommand{\DOIprefix}{doi:}
\providecommand{\ArXivprefix}{arXiv:}
\providecommand{\URLprefix}{URL: }
\providecommand{\Pubmedprefix}{pmid:}
\providecommand{\doi}[1]{\href{http://dx.doi.org/#1}{\path{#1}}}
\providecommand{\Pubmed}[1]{\href{pmid:#1}{\path{#1}}}
\providecommand{\bibinfo}[2]{#2}
\ifx\xfnm\relax \def\xfnm[#1]{\unskip,\space#1}\fi
\bibitem[{{Abbott} et~al.(2022){Abbott}, {Aguena}, {Alarcon}, et~al. and {DES
  Collaboration}}]{DES3year}
\bibinfo{author}{{Abbott}, T.M.C.}, \bibinfo{author}{{Aguena}, M.},
  \bibinfo{author}{{Alarcon}, A.}, \bibinfo{author}{et~al.},
  \bibinfo{author}{{DES Collaboration}}, \bibinfo{year}{2022}.
\newblock \bibinfo{title}{{Dark Energy Survey Year 3 results: Cosmological
  constraints from galaxy clustering and weak lensing}}.
\newblock \bibinfo{journal}{Physical Review D} \bibinfo{volume}{105},
  \bibinfo{pages}{023520}.
\newblock \DOIprefix\doi{10.1103/PhysRevD.105.023520},
  \href{http://arxiv.org/abs/2105.13549}{{\tt arXiv:2105.13549}}.
\bibitem[{Alam et~al.(2017)Alam, Ata, Bailey, Beutler and
  et~al.}]{10.1093/mnras/stx721}
\bibinfo{author}{Alam, S.}, \bibinfo{author}{Ata, M.}, \bibinfo{author}{Bailey,
  S.}, \bibinfo{author}{Beutler, F.}, \bibinfo{author}{et~al.},
  \bibinfo{year}{2017}.
\newblock \bibinfo{title}{{The clustering of galaxies in the completed SDSS-III
  Baryon Oscillation Spectroscopic Survey: cosmological analysis of the DR12
  galaxy sample}}.
\newblock \bibinfo{journal}{Monthly Notices of the Royal Astronomical Society}
  \bibinfo{volume}{470}, \bibinfo{pages}{2617--2652}.
\newblock \DOIprefix\doi{10.1093/mnras/stx721}.
\bibitem[{Alam et~al.(2021)Alam, Aubert, Avila, Balland,  and
  et~al.}]{PhysRevD.103.083533}
\bibinfo{author}{Alam, S.}, \bibinfo{author}{Aubert, M.},
  \bibinfo{author}{Avila, S.}, \bibinfo{author}{Balland, C.}, ,
  \bibinfo{author}{et~al.}, \bibinfo{year}{2021}.
\newblock \bibinfo{title}{Completed sdss-iv extended baryon oscillation
  spectroscopic survey: Cosmological implications from two decades of
  spectroscopic surveys at the apache point observatory}.
\newblock \bibinfo{journal}{Phys. Rev. D} \bibinfo{volume}{103},
  \bibinfo{pages}{083533}.
\newblock \URLprefix
  \url{https://link.aps.org/doi/10.1103/PhysRevD.103.083533},
  \DOIprefix\doi{10.1103/PhysRevD.103.083533}.
\bibitem[{Audren et~al.(2013)Audren, Lesgourgues, Benabed and
  Prunet}]{Audren:2012wb}
\bibinfo{author}{Audren, B.}, \bibinfo{author}{Lesgourgues, J.},
  \bibinfo{author}{Benabed, K.}, \bibinfo{author}{Prunet, S.},
  \bibinfo{year}{2013}.
\newblock \bibinfo{title}{{Conservative Constraints on Early Cosmology: an
  illustration of the Monte Python cosmological parameter inference code}}.
\newblock \bibinfo{journal}{JCAP} \bibinfo{volume}{1302}, \bibinfo{pages}{001}.
\newblock \DOIprefix\doi{10.1088/1475-7516/2013/02/001},
  \href{http://arxiv.org/abs/1210.7183}{{\tt arXiv:1210.7183}}.
\bibitem[{Bautista et~al.(2020)Bautista, Paviot, Vargas Magaña, de la Torre
  and et~al.}]{10.1093/mnras/staa2800}
\bibinfo{author}{Bautista, J.E.}, \bibinfo{author}{Paviot, R.},
  \bibinfo{author}{Vargas Magaña, M.}, \bibinfo{author}{de la Torre, S.},
  \bibinfo{author}{et~al.}, \bibinfo{year}{2020}.
\newblock \bibinfo{title}{{The completed SDSS-IV extended Baryon Oscillation
  Spectroscopic Survey: measurement of the BAO and growth rate of structure of
  the luminous red galaxy sample from the anisotropic correlation function
  between redshifts 0.6 and 1}}.
\newblock \bibinfo{journal}{Monthly Notices of the Royal Astronomical Society}
  \bibinfo{volume}{500}, \bibinfo{pages}{736--762}.
\newblock \DOIprefix\doi{10.1093/mnras/staa2800}.
\bibitem[{Blas et~al.(2011)Blas, Lesgourgues and Tram}]{DiegoBlas_2011}
\bibinfo{author}{Blas, D.}, \bibinfo{author}{Lesgourgues, J.},
  \bibinfo{author}{Tram, T.}, \bibinfo{year}{2011}.
\newblock \bibinfo{title}{The cosmic linear anisotropy solving system (class).
  part ii: Approximation schemes}.
\newblock \bibinfo{journal}{Journal of Cosmology and Astroparticle Physics}
  \bibinfo{volume}{2011}, \bibinfo{pages}{034}.
\newblock \URLprefix \url{https://dx.doi.org/10.1088/1475-7516/2011/07/034},
  \DOIprefix\doi{10.1088/1475-7516/2011/07/034}.
\bibitem[{du~Mas~des Bourboux et~al.(2020)du~Mas~des Bourboux, Rich,
  Font-Ribera, de~Sainte~Agathe and et~al.}]{du_Mas_des_Bourboux_2020}
\bibinfo{author}{du~Mas~des Bourboux, H.}, \bibinfo{author}{Rich, J.},
  \bibinfo{author}{Font-Ribera, A.}, \bibinfo{author}{de~Sainte~Agathe, V.},
  \bibinfo{author}{et~al.}, \bibinfo{year}{2020}.
\newblock \bibinfo{title}{The completed sdss-iv extended baryon oscillation
  spectroscopic survey: Baryon acoustic oscillations with ly$\alpha$ forests}.
\newblock \bibinfo{journal}{The Astrophysical Journal} \bibinfo{volume}{901},
  \bibinfo{pages}{153}.
\newblock \URLprefix \url{https://dx.doi.org/10.3847/1538-4357/abb085},
  \DOIprefix\doi{10.3847/1538-4357/abb085}.
\bibitem[{Brinckmann and Lesgourgues(2019)}]{BRINCKMANN2019100260}
\bibinfo{author}{Brinckmann, T.}, \bibinfo{author}{Lesgourgues, J.},
  \bibinfo{year}{2019}.
\newblock \bibinfo{title}{Montepython 3: Boosted mcmc sampler and other
  features}.
\newblock \bibinfo{journal}{Physics of the Dark Universe} \bibinfo{volume}{24},
  \bibinfo{pages}{100260}.
\newblock \DOIprefix\doi{https://doi.org/10.1016/j.dark.2018.100260}.
\bibitem[{{DESI Collaboration} et~al.(2024){DESI Collaboration}, {Adame},
  {Aguilar}, {Ahlen} and {et al}.}]{2024arXiv240403002D}
\bibinfo{author}{{DESI Collaboration}}, \bibinfo{author}{{Adame}, A.G.},
  \bibinfo{author}{{Aguilar}, J.}, \bibinfo{author}{{Ahlen}, S.},
  \bibinfo{author}{{et al}.}, \bibinfo{year}{2024}.
\newblock \bibinfo{title}{{DESI 2024 VI: Cosmological Constraints from the
  Measurements of Baryon Acoustic Oscillations}}.
\newblock \bibinfo{journal}{arXiv e-prints} ,
  \bibinfo{pages}{arXiv:2404.03002}\DOIprefix\doi{10.48550/arXiv.2404.03002},
  \href{http://arxiv.org/abs/2404.03002}{{\tt arXiv:2404.03002}}.
\bibitem[{{Freedman}(2021)}]{2021ApJ...919...16F}
\bibinfo{author}{{Freedman}, W.L.}, \bibinfo{year}{2021}.
\newblock \bibinfo{title}{{Measurements of the Hubble Constant: Tensions in
  Perspective}}.
\newblock \bibinfo{journal}{The Astrophysical Journal} \bibinfo{volume}{919},
  \bibinfo{pages}{16}.
\newblock \DOIprefix\doi{10.3847/1538-4357/ac0e95},
  \href{http://arxiv.org/abs/2106.15656}{{\tt arXiv:2106.15656}}.
\bibitem[{Gil-Marín et~al.(2020)Gil-Marín, Bautista, Paviot, Vargas-Magaña
  and et~al.}]{10.1093/mnras/staa2455}
\bibinfo{author}{Gil-Marín, H.}, \bibinfo{author}{Bautista, J.E.},
  \bibinfo{author}{Paviot, R.}, \bibinfo{author}{Vargas-Magaña, M.},
  \bibinfo{author}{et~al.}, \bibinfo{year}{2020}.
\newblock \bibinfo{title}{{The Completed SDSS-IV extended Baryon Oscillation
  Spectroscopic Survey: measurement of the BAO and growth rate of structure of
  the luminous red galaxy sample from the anisotropic power spectrum between
  redshifts 0.6 and 1.0}}.
\newblock \bibinfo{journal}{Monthly Notices of the Royal Astronomical Society}
  \bibinfo{volume}{498}, \bibinfo{pages}{2492--2531}.
\newblock \URLprefix \url{https://doi.org/10.1093/mnras/staa2455},
  \DOIprefix\doi{10.1093/mnras/staa2455}.
\bibitem[{G\'omez-Valent(2022)}]{Gomez-Valent:2022hkb}
\bibinfo{author}{G\'omez-Valent, A.}, \bibinfo{year}{2022}.
\newblock \bibinfo{title}{{Fast test to assess the impact of marginalization in
  Monte~Carlo analyses and its application to cosmology}}.
\newblock \bibinfo{journal}{Phys. Rev. D} \bibinfo{volume}{106},
  \bibinfo{pages}{063506}.
\newblock \DOIprefix\doi{10.1103/PhysRevD.106.063506},
  \href{http://arxiv.org/abs/2203.16285}{{\tt arXiv:2203.16285}}.
\bibitem[{{Heymans} et~al.(2021){Heymans}, {Tr{\"o}ster}, {Asgari} and
  et~al.}]{Kids}
\bibinfo{author}{{Heymans}, C.}, \bibinfo{author}{{Tr{\"o}ster}, T.},
  \bibinfo{author}{{Asgari}, M.}, \bibinfo{author}{et~al.},
  \bibinfo{year}{2021}.
\newblock \bibinfo{title}{{KiDS-1000 Cosmology: Multi-probe weak gravitational
  lensing and spectroscopic galaxy clustering constraints}}.
\newblock \bibinfo{journal}{Astronomy and Astrophysics} \bibinfo{volume}{646},
  \bibinfo{pages}{A140}.
\newblock \DOIprefix\doi{10.1051/0004-6361/202039063},
  \href{http://arxiv.org/abs/2007.15632}{{\tt arXiv:2007.15632}}.
\bibitem[{Hou et~al.(2020)Hou, Sánchez, Ross, Smith and
  et~al.}]{10.1093/mnras/staa3234}
\bibinfo{author}{Hou, J.}, \bibinfo{author}{Sánchez, A.G.},
  \bibinfo{author}{Ross, A.J.}, \bibinfo{author}{Smith, A.},
  \bibinfo{author}{et~al.}, \bibinfo{year}{2020}.
\newblock \bibinfo{title}{{The completed SDSS-IV extended Baryon Oscillation
  Spectroscopic Survey: BAO and RSD measurements from anisotropic clustering
  analysis of the quasar sample in configuration space between redshift 0.8 and
  2.2}}.
\newblock \bibinfo{journal}{Monthly Notices of the Royal Astronomical Society}
  \bibinfo{volume}{500}, \bibinfo{pages}{1201--1221}.
\newblock \DOIprefix\doi{10.1093/mnras/staa3234}.
\bibitem[{Howlett et~al.(2015)Howlett, Ross, Samushia, Percival and
  Manera}]{10.1093/mnras/stu2693}
\bibinfo{author}{Howlett, C.}, \bibinfo{author}{Ross, A.J.},
  \bibinfo{author}{Samushia, L.}, \bibinfo{author}{Percival, W.J.},
  \bibinfo{author}{Manera, M.}, \bibinfo{year}{2015}.
\newblock \bibinfo{title}{{The clustering of the SDSS main galaxy sample –
  II. Mock galaxy catalogues and a measurement of the growth of structure from
  redshift space distortions at z = 0.15}}.
\newblock \bibinfo{journal}{Monthly Notices of the Royal Astronomical Society}
  \bibinfo{volume}{449}, \bibinfo{pages}{848--866}.
\newblock \DOIprefix\doi{10.1093/mnras/stu2693}.
\bibitem[{{Lemos} et~al.(2021){Lemos}, {Raveri}, {Campos}, et~al. and {DES
  Collaboration}}]{2021MNRAS.505.6179L}
\bibinfo{author}{{Lemos}, P.}, \bibinfo{author}{{Raveri}, M.},
  \bibinfo{author}{{Campos}, A.}, \bibinfo{author}{et~al.},
  \bibinfo{author}{{DES Collaboration}}, \bibinfo{year}{2021}.
\newblock \bibinfo{title}{{Assessing tension metrics with dark energy survey
  and Planck data}}.
\newblock \bibinfo{journal}{Monthly Notices of the Royal Astronomical Society}
  \bibinfo{volume}{505}, \bibinfo{pages}{6179--6194}.
\newblock \DOIprefix\doi{10.1093/mnras/stab1670},
  \href{http://arxiv.org/abs/2012.09554}{{\tt arXiv:2012.09554}}.
\bibitem[{Moresco(2015)}]{10.1093/mnrasl/slv037}
\bibinfo{author}{Moresco, M.}, \bibinfo{year}{2015}.
\newblock \bibinfo{title}{{Raising the bar: new constraints on the Hubble
  parameter with cosmic chronometers at z ~ 2}}.
\newblock \bibinfo{journal}{Monthly Notices of the Royal Astronomical Society:
  Letters} \bibinfo{volume}{450}, \bibinfo{pages}{L16--L20}.
\newblock \DOIprefix\doi{10.1093/mnrasl/slv037}.
\bibitem[{Moresco(2023)}]{Moresco:2023zys}
\bibinfo{author}{Moresco, M.}, \bibinfo{year}{2023}.
\newblock \bibinfo{title}{{Addressing the Hubble tension with cosmic
  chronometers}} \href{http://arxiv.org/abs/2307.09501}{{\tt
  arXiv:2307.09501}}.
\bibitem[{{Moresco} et~al.(2012){Moresco}, {Cimatti}, {Jimenez} and
  et~al.}]{moresco12}
\bibinfo{author}{{Moresco}, M.}, \bibinfo{author}{{Cimatti}, A.},
  \bibinfo{author}{{Jimenez}, R.}, \bibinfo{author}{et~al.},
  \bibinfo{year}{2012}.
\newblock \bibinfo{title}{{Improved constraints on the expansion rate of the
  Universe up to z \~{} 1.1 from the spectroscopic evolution of cosmic
  chronometers}}.
\newblock \bibinfo{journal}{Journal of Cosmology and Astroparticle Physics}
  \bibinfo{volume}{8}, \bibinfo{pages}{006}.
\newblock \DOIprefix\doi{10.1088/1475-7516/2012/08/006},
  \href{http://arxiv.org/abs/1201.3609}{{\tt arXiv:1201.3609}}.
\bibitem[{{Moresco} et~al.(2016){Moresco}, {Pozzetti}, {Cimatti}, {Jimenez},
  {Maraston}, {Verde}, {Thomas}, {Citro}, {Tojeiro} and {Wilkinson}}]{CC2}
\bibinfo{author}{{Moresco}, M.}, \bibinfo{author}{{Pozzetti}, L.},
  \bibinfo{author}{{Cimatti}, A.}, \bibinfo{author}{{Jimenez}, R.},
  \bibinfo{author}{{Maraston}, C.}, \bibinfo{author}{{Verde}, L.},
  \bibinfo{author}{{Thomas}, D.}, \bibinfo{author}{{Citro}, A.},
  \bibinfo{author}{{Tojeiro}, R.}, \bibinfo{author}{{Wilkinson}, D.},
  \bibinfo{year}{2016}.
\newblock \bibinfo{title}{{A 6\% measurement of the Hubble parameter at
  z\~{}0.45: direct evidence of the epoch of cosmic re-acceleration}}.
\newblock \bibinfo{journal}{Journal of Cosmology and Astroparticle Physics}
  \bibinfo{volume}{5}, \bibinfo{pages}{014}.
\newblock \DOIprefix\doi{10.1088/1475-7516/2016/05/014},
  \href{http://arxiv.org/abs/1601.01701}{{\tt arXiv:1601.01701}}.
\bibitem[{Neveux et~al.(2020)Neveux, Burtin, de Mattia, Smith and
  et~al.}]{10.1093/mnras/staa2780}
\bibinfo{author}{Neveux, R.}, \bibinfo{author}{Burtin, E.},
  \bibinfo{author}{de Mattia, A.}, \bibinfo{author}{Smith, A.},
  \bibinfo{author}{et~al.}, \bibinfo{year}{2020}.
\newblock \bibinfo{title}{{The completed SDSS-IV extended Baryon Oscillation
  Spectroscopic Survey: BAO and RSD measurements from the anisotropic power
  spectrum of the quasar sample between redshift 0.8 and 2.2}}.
\newblock \bibinfo{journal}{Monthly Notices of the Royal Astronomical Society}
  \bibinfo{volume}{499}, \bibinfo{pages}{210--229}.
\newblock \DOIprefix\doi{10.1093/mnras/staa2780}.
\bibitem[{{Planck Collaboration} et~al.(2020){Planck Collaboration}, {Aghanim},
  {Akrami}, {Ashdown} and et~al.}]{Planckcosmo2018}
\bibinfo{author}{{Planck Collaboration}}, \bibinfo{author}{{Aghanim}, N.},
  \bibinfo{author}{{Akrami}, Y.}, \bibinfo{author}{{Ashdown}, M.},
  \bibinfo{author}{et~al.}, \bibinfo{year}{2020}.
\newblock \bibinfo{title}{{Planck 2018 results. VI. Cosmological parameters}}.
\newblock \bibinfo{journal}{Astronomy and Astrophysics} \bibinfo{volume}{641},
  \bibinfo{pages}{A6}.
\newblock \DOIprefix\doi{10.1051/0004-6361/201833910},
  \href{http://arxiv.org/abs/1807.06209}{{\tt arXiv:1807.06209}}.
\bibitem[{Raveri and Doux(2021)}]{PhysRevD.104.043504}
\bibinfo{author}{Raveri, M.}, \bibinfo{author}{Doux, C.}, \bibinfo{year}{2021}.
\newblock \bibinfo{title}{Non-gaussian estimates of tensions in cosmological
  parameters}.
\newblock \bibinfo{journal}{Phys. Rev. D} \bibinfo{volume}{104},
  \bibinfo{pages}{043504}.
\newblock \URLprefix
  \url{https://link.aps.org/doi/10.1103/PhysRevD.104.043504},
  \DOIprefix\doi{10.1103/PhysRevD.104.043504}.
\bibitem[{Raveri and Hu(2019)}]{PhysRevD.99.043506}
\bibinfo{author}{Raveri, M.}, \bibinfo{author}{Hu, W.}, \bibinfo{year}{2019}.
\newblock \bibinfo{title}{Concordance and discordance in cosmology}.
\newblock \bibinfo{journal}{Phys. Rev. D} \bibinfo{volume}{99},
  \bibinfo{pages}{043506}.
\newblock \URLprefix \url{https://link.aps.org/doi/10.1103/PhysRevD.99.043506},
  \DOIprefix\doi{10.1103/PhysRevD.99.043506}.
\bibitem[{Raveri et~al.(2020)Raveri, Zacharegkas and Hu}]{PhysRevD.101.103527}
\bibinfo{author}{Raveri, M.}, \bibinfo{author}{Zacharegkas, G.},
  \bibinfo{author}{Hu, W.}, \bibinfo{year}{2020}.
\newblock \bibinfo{title}{Quantifying concordance of correlated cosmological
  data sets}.
\newblock \bibinfo{journal}{Phys. Rev. D} \bibinfo{volume}{101},
  \bibinfo{pages}{103527}.
\newblock \URLprefix
  \url{https://link.aps.org/doi/10.1103/PhysRevD.101.103527},
  \DOIprefix\doi{10.1103/PhysRevD.101.103527}.
\bibitem[{{Riess} and {Breuval}(2023)}]{2023arXiv230810954R}
\bibinfo{author}{{Riess}, A.G.}, \bibinfo{author}{{Breuval}, L.},
  \bibinfo{year}{2023}.
\newblock \bibinfo{title}{{The Local Value of H$_0$}}.
\newblock \bibinfo{journal}{arXiv e-prints} ,
  \bibinfo{pages}{arXiv:2308.10954}\DOIprefix\doi{10.48550/arXiv.2308.10954},
  \href{http://arxiv.org/abs/2308.10954}{{\tt arXiv:2308.10954}}.
\bibitem[{{Riess} et~al.(2022){Riess}, {Yuan}, {Macri} and et~al.}]{Riess2022}
\bibinfo{author}{{Riess}, A.G.}, \bibinfo{author}{{Yuan}, W.},
  \bibinfo{author}{{Macri}, L.M.}, \bibinfo{author}{et~al.},
  \bibinfo{year}{2022}.
\newblock \bibinfo{title}{{A Comprehensive Measurement of the Local Value of
  the Hubble Constant with 1 km s$^{-1}$ Mpc$^{-1}$ Uncertainty from the Hubble
  Space Telescope and the SH0ES Team}}.
\newblock \bibinfo{journal}{The Astrophysical Journal Letters}
  \bibinfo{volume}{934}, \bibinfo{pages}{L7}.
\newblock \DOIprefix\doi{10.3847/2041-8213/ac5c5b},
  \href{http://arxiv.org/abs/2112.04510}{{\tt arXiv:2112.04510}}.
\bibitem[{{Sch{\"o}neberg} et~al.(2022){Sch{\"o}neberg}, {Abell{\'a}n},
  {S{\'a}nchez}, {Witte}, {Poulin} and {Lesgourgues}}]{2022PhR...984....1S}
\bibinfo{author}{{Sch{\"o}neberg}, N.}, \bibinfo{author}{{Abell{\'a}n}, G.F.},
  \bibinfo{author}{{S{\'a}nchez}, A.P.}, \bibinfo{author}{{Witte}, S.J.},
  \bibinfo{author}{{Poulin}, V.}, \bibinfo{author}{{Lesgourgues}, J.},
  \bibinfo{year}{2022}.
\newblock \bibinfo{title}{{The H$_{0}$ Olympics: A fair ranking of proposed
  models}}.
\newblock \bibinfo{journal}{Physics Reports} \bibinfo{volume}{984},
  \bibinfo{pages}{1--55}.
\newblock \DOIprefix\doi{10.1016/j.physrep.2022.07.001},
  \href{http://arxiv.org/abs/2107.10291}{{\tt arXiv:2107.10291}}.
\bibitem[{Scolnic et~al.(2022)}]{Scolnic:2021amr}
\bibinfo{author}{Scolnic, D.}, et~al., \bibinfo{year}{2022}.
\newblock \bibinfo{title}{{The Pantheon+ Analysis: The Full Data Set and
  Light-curve Release}}.
\newblock \bibinfo{journal}{Astrophys. J.} \bibinfo{volume}{938},
  \bibinfo{pages}{113}.
\newblock \DOIprefix\doi{10.3847/1538-4357/ac8b7a},
  \href{http://arxiv.org/abs/2112.03863}{{\tt arXiv:2112.03863}}.
\bibitem[{{Simon} et~al.(2005){Simon}, {Verde} and {Jimenez}}]{simon05}
\bibinfo{author}{{Simon}, J.}, \bibinfo{author}{{Verde}, L.},
  \bibinfo{author}{{Jimenez}, R.}, \bibinfo{year}{2005}.
\newblock \bibinfo{title}{{Constraints on the redshift dependence of the dark
  energy potential}}.
\newblock \bibinfo{journal}{Physical Review D} \bibinfo{volume}{71},
  \bibinfo{pages}{123001}.
\newblock \DOIprefix\doi{10.1103/PhysRevD.71.123001},
  \href{http://arxiv.org/abs/astro-ph/0412269}{{\tt arXiv:astro-ph/0412269}}.
\bibitem[{{Stern} et~al.(2010){Stern}, {Jimenez}, {Verde}, {Kamionkowski} and
  {Stanford}}]{stern10}
\bibinfo{author}{{Stern}, D.}, \bibinfo{author}{{Jimenez}, R.},
  \bibinfo{author}{{Verde}, L.}, \bibinfo{author}{{Kamionkowski}, M.},
  \bibinfo{author}{{Stanford}, S.A.}, \bibinfo{year}{2010}.
\newblock \bibinfo{title}{{Cosmic chronometers: constraining the equation of
  state of dark energy. I: H(z) measurements}}.
\newblock \bibinfo{journal}{Journal of Cosmology and Astroparticle Physics}
  \bibinfo{volume}{2}, \bibinfo{pages}{008}.
\newblock \DOIprefix\doi{10.1088/1475-7516/2010/02/008},
  \href{http://arxiv.org/abs/0907.3149}{{\tt arXiv:0907.3149}}.
\bibitem[{Vagnozzi(2023)}]{Vagnozzi_2023}
\bibinfo{author}{Vagnozzi, S.}, \bibinfo{year}{2023}.
\newblock \bibinfo{title}{Seven hints that early-time new physics alone is not
  sufficient to solve the hubble tension}.
\newblock \bibinfo{journal}{Universe} \bibinfo{volume}{9},
  \bibinfo{pages}{393}.
\newblock \URLprefix \url{http://dx.doi.org/10.3390/universe9090393},
  \DOIprefix\doi{10.3390/universe9090393}.
\bibitem[{{Zhang} et~al.(2014){Zhang}, {Zhang}, {Yuan}, {Liu}, {Zhang} and
  {Sun}}]{zhang14}
\bibinfo{author}{{Zhang}, C.}, \bibinfo{author}{{Zhang}, H.},
  \bibinfo{author}{{Yuan}, S.}, \bibinfo{author}{{Liu}, S.},
  \bibinfo{author}{{Zhang}, T.J.}, \bibinfo{author}{{Sun}, Y.C.},
  \bibinfo{year}{2014}.
\newblock \bibinfo{title}{{Four new observational H(z) data from luminous red
  galaxies in the Sloan Digital Sky Survey data release seven}}.
\newblock \bibinfo{journal}{Research in Astronomy and Astrophysics}
  \bibinfo{volume}{14}, \bibinfo{pages}{1221--1233}.
\newblock \DOIprefix\doi{10.1088/1674-4527/14/10/002},
  \href{http://arxiv.org/abs/1207.4541}{{\tt arXiv:1207.4541}}.

\end{thebibliography}

\end{document}